\documentclass[aps,prl,reprint,superscriptaddress,amssymb,floatfix,showpacs]{revtex4-1}
\usepackage{amsmath}
\usepackage{amsfonts}
\usepackage{verbatim}
\usepackage{amsthm}
\usepackage{amssymb}
\usepackage{graphicx}
\usepackage{hyperref}

\newcommand{\fig}{Fig.\ }

\newcommand{\eqn}{Eq.\ }

\newcommand{\alphav}[1]{\langle w_{#1} \rangle}
\newcommand{\alphavz}{\langle w \rangle}

\begin{document}


\title{Emergent motion of condensates in mass-transport models}

\author{Ori Hirschberg}
\author{David Mukamel}
\affiliation{Department of Physics of Complex Systems, Weizmann
Institute of Science, 76100 Rehovot, Israel}
\author{Gunter M. Sch{\"u}tz}
\affiliation{Institut f\"ur Festk\"orperforschung,
Forschungszentrum J\"ulich, 52425 J\"ulich, Germany}

\date{\today}



\begin{abstract}
We examine the effect of spatial correlations on the phenomenon of
real-space condensation in driven mass-transport systems. We suggest
that in a broad class of models with a spatially correlated steady
state, the condensate drifts with a non-vanishing velocity. We
present a robust mechanism leading to this condensate drift. This is
done within the framework of a generalized zero-range process (ZRP)
in which, unlike the usual ZRP, the steady state is not a product
measure. The validity of the mechanism in other mass-transport
models is discussed.
\end{abstract}


\maketitle

Nonequilibrium condensation, whereby a macroscopic fraction of
microscopic constituents of a system accumulates in a local region,
is a common feature of many mass-transport systems. Examples include
shaken granular gasses \cite{vanderweele2001}, vehicular traffic
\cite{BusRouteModel1998,Chowdhury2000TrafficReview,kaupuzsetal2005zrptraffic},
the macroeconomics of wealth distribution
\cite{BouchaudMezard2000WealthCondensation,BurdaEtal2002WealthZRP},
and others \cite{MajumdarEtalChipping1998,networksevolutionbook}.
Mechanisms which can lead to the formation of condensates have been
studied extensively in recent years, mainly by analyzing
prototypical toy models. A primary role in these studies was played
by the zero-range process (ZRP), an exactly-solvable model in which
particles hop between sites with rates which depend only on the
number of particles in the departure site
\cite{evanszrpreview,MajumdarCondensationLesHouches,Schadschneider2010Book}.
Extensions and variations of the ZRP have been used to study the
emergence of multiple condensates
\cite{Schwarzkopf2008MultipleConds}, first order condensation
transitions \cite{GrosskinskySchutzZRP1stOrder,ChlebounThesis} and
the effect of interactions \cite{EvansEtal2006PairFactorized} and
disorder \cite{GodrecheLuck2012Inhomogeneous} on condensation.
Moreover, one-dimensional phase separation transitions in exclusion
processes and other driven diffusive systems can quite generally be
understood by a mapping on ZRPs \cite{kafrietal2002criterion}.

The dynamics of condensates is less well explored. In the ZRP, where
condensation takes place when a macroscopic fraction of particles
occupies a single site, the resulting condensate does not drift in
the thermodynamic limit
\cite{godrecheluck2005condensate,beltramlandim2008,Landim2012AsymmetricZRPCondensate,ChlebounThesis}.
It is shown below that this is related to the fact that the steady
state of the ZRP is a product measure. In some real-world systems,
however, condensates are in continual motion. For example, traffic
jams, which can be viewed as condensates, are known to propagate
along congested roads
\cite{LighthillWitham1955,Treiterer1975Traffic,Schadschneider2010Book}.
Recently, two variants of the ZRP were also found to relax to a time
dependent state in which the condensate performs a drift motion: one
is a ZRP with non-Markovian hopping rates
\cite{HirschbergEtal2009,HirschbergEtal2012Long}, and the other is a
model with ``explosive condensation''
\cite{BartekEvans2012Explosive}. To date, there is no systematic
understanding of the mechanism by which a macroscopic condensate
motion emerges from the underlying nonequilibrium microscopic
dynamics.

In this Letter, we study how spatial correlations in the steady
states may lead the condensate to drift with a non-vanishing
velocity. We do so by introducing a generalization of the ZRP whose
steady state does not factorize. Within its framework, we identify
the mechanism which generates the drift. The analysis is based on
numeric simulations and on a mean-field approximation which captures
the essential effect of correlations in the condensed phase, and
thus elucidates the different observed modes of condensate motion.
The drift mechanism which we identify is robust and therefore it is
expected to be valid in a broad class of spatially-correlated
mass-transport systems.

We focus on a class of stochastic one-dimensional models defined on
a ring with $L$ sites. At any given time, each site $i$ is occupied
by $n_i$ particles with $\sum_i n_i=N$, $n_i \geq 0$. The model
evolves by a totally asymmetric hopping process whereby particles
hop from site $i$ to $i+1$ with a rate which depends on the
occupation numbers $n_i$ and $n_{i-1}$. This is a generalization of
the usual ZRP in which the rate depends only on $n_i$. More
specifically, we choose the hopping rates to be of the form
\begin{equation}\label{eq:modeldscrp}
n_{i-1},n_i,n_{i+1} \xrightarrow{w(n_{i-1})u(n_i)}
n_{i-1},n_i-1,n_{i+1}+1,
\end{equation}
with rates
\begin{equation}\label{eq:hoprates}
u(n_i) = 1 + \frac{b}{n_i}, \qquad w (n_{i-1}) = \left\{
\begin{array}{ll} 1 & n_{i-1} \neq 0 \\
\alpha & n_{i-1} = 0
\end{array} \right..
\end{equation}
The particular form of $u(n)$ is motivated by the fact that in the
usual ZRP, which corresponds to $\alpha = 1$, this choice with $b>2$
leads to a condensation transition \cite{EvansZRPcondensation}. The
rate $w$ with $\alpha \neq 1$ represents an interaction between
nearest-neighbor sites. According the dynamical rules
(\ref{eq:modeldscrp})--(\ref{eq:hoprates}), at every short time
interval $dt$, each site $i$ whose occupation $n_i \geq 1$ may eject
a particle with a probability $u(n_i)dt$, as long as the preceding
site ($i-1$) is occupied. If the preceding site is empty, this
probability changes to $\alpha u(n_i)dt$. The model has three
parameters: $b$, $\alpha$, and the density $\rho \equiv N/L$ which
is conserved by the dynamics.

\begin{figure*}
        \includegraphics[width=0.99\textwidth]{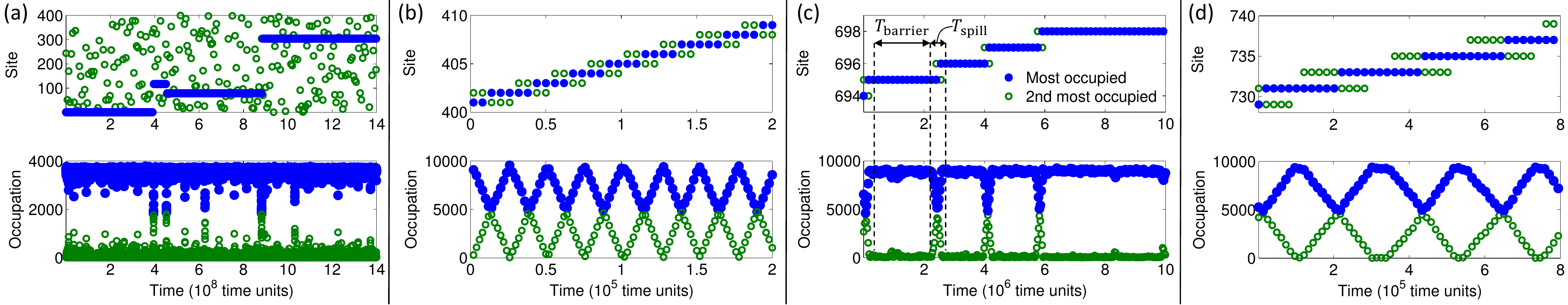}
        \caption{The location (top) and occupation (bottom) of the most occupied
        (\textbullet) and 2$^\text{nd}$ most occupied ($\circ$)
        sites for several values of $\alpha$. (a) $\alpha = 1$,
        i.e., the usual ZRP. The condensate is stable for long times,
        and relocates to random distant sites. (b) $\alpha = 1.5$.
        The condensate advances through a slinky motion from one site
        to the next. (c) $\alpha = 1.05$. The condensate remains on
        each site for a long time before ``spilling'' to the next.
        The definitions of $T_\text{barrier}$ and $T_\text{spill}$ are
        indicated. (d) $\alpha = 0.5$. The condensate skips every
        other site. In all cases $b = 3$, $\rho = 10$ and
        $L=1000$, except (a) where $L=400$. Note the different time
        scales.}
        \label{fig:CondDrift}
\end{figure*}

In the usual ZRP (the case of $\alpha = 1$), the stationary
distribution is known to factorize into a product of single site
terms, and so can be exactly calculated
\cite{Spitzer1970,andjel1982,evanszrpreview}. This factorization
property renders the ZRP quite special, as slight variations of the
ZRP dynamics result in non-factorizable models. To probe the effect
of spatial correlations on the condensate, we choose for simplicity
$w$ to be of the form (\ref{eq:hoprates}), which leads to a
spatially correlated steady-state when $\alpha \neq 1$. The drift
motion which we describe below is also found for other forms of
$w(n)$, such as cases where $w(n) \neq 1$ for finitely many values
of $n$, and for other choices of $u(n)$ which give rise to
condensation \cite{FutureLongPaper}.


We have carried out Monte Carlo simulations of the dynamics
(\ref{eq:modeldscrp})--(\ref{eq:hoprates}) for several values of
$\alpha$ in a system of size $L=1000$ and density $\rho = 10$. After
the system has relaxed to its steady state, the dynamics of the
condensate was examined by tracking the position of the most
occupied sites over time. The results are presented in \fig
\ref{fig:CondDrift} and in videos in the supplemental material
\cite{supplemental}. In the usual asymmetric ZRP (\fig
\ref{fig:CondDrift}a), it is known that the condensate is static up
to timescales of order $L^b$ and then it relocates to a random site
due to fluctuations
\cite{godrecheluck2005condensate,beltramlandim2008,Landim2012AsymmetricZRPCondensate,ChlebounThesis}.
There is a striking qualitative difference in the dynamics of model
(\ref{eq:modeldscrp})--(\ref{eq:hoprates}) when $\alpha \neq 1$
(\fig \ref{fig:CondDrift}b--\ref{fig:CondDrift}d), where the
condensate is clearly seen to drift along the lattice. The
condensate is seen to move from one site to the next when $\alpha >
1$ (\fig \ref{fig:CondDrift}b), or to skip every other site when
$\alpha < 1$ (\fig \ref{fig:CondDrift}d). In both cases, when
$\alpha$ is not too close to 1 the motion is ``slinky''-like, with
the condensate ``spilling'' from an old site to a new one
immediately after the previous spilling has completed. The drift
becomes somewhat less regular when $\alpha$ is close to 1. In this
regime, the slinky motion is interrupted by periods of time when the
condensate occupies a single site, before the spilling process is
initiated (\fig \ref{fig:CondDrift}c). This, however, is argued
below to be a crossover mode, and the interval in $\alpha$ in which
it is observed shrinks in the large $L$ limit.

To understand these results we propose a mean-field analysis of the
model in which the occupations of all sites are considered
independent, but might not be identically distributed. Within this
approximation, the current which arrives into site $i$ from site
$i-1$ is a Poisson process whose rate we denote by $J_i$. The
probability $P_i(n_i)$ to find $n_i$ particles in site $i$ thus
evolves according to
\begin{align}\label{eq:mastereq}
\frac{dP_i(n_i)}{dt} = {}& P_i(n_i-1)J_i +
P_i(n_i+1)\alphav{i}u(n_i+1)
\nonumber \\
&{} - P_i(n_i)\bigl(J_i + \alphav{i} u(n_i)\bigr),
\end{align}
where
\begin{equation}\label{eq:betadef}
\alphav{i} \equiv \sum_{n=0}^\infty P_{i-1}(n) w(n) = 1 +
(\alpha-1)P_{i-1}(0)
\end{equation}
encodes the mean effect of site $i-1$ on the hopping rate out of
site $i$. Equation (\ref{eq:mastereq}) is valid also when $n_i = 0$
with the definition $P_i(-1) \equiv 0$. Equations
(\ref{eq:mastereq}) and (\ref{eq:betadef}) are to be solved with the
self consistency condition $J_{i+1} = \sum_n P_i(n)\alphav{i} u(n)$.

At low density, the system is in a subcritical, disordered phase
(this will be shown below). In this homogeneous phase,  $P_i(n) =
P(n)$ and $J_i = J$ for all sites $i$. At higher densities, however,
condensation takes place, where the translational symmetry is
spontaneously broken and both $P_i(n)$ and $J_i$ depend on the
distance of site $i$ from the condensate. This dependence of $P$ and
$J$ on $i$ is a result of the correlations which exist in the steady
state of the model, and it provides the mechanism for the condensate
drift: a nonhomogeneous $J_i$ implies that in some sites the
outflowing current is smaller than the incoming current, leading
these sites to accumulate particles while other sites are similarly
being depleted of particles. We shall now demonstrate that this
occurs in our model.

In the homogeneous (subcritical and critical) phases, the model
eventually reaches a steady state. In the nonhomogeneous
supercritical phase, however, the condensate location keeps moving
with time. The analysis of this time-dependent phase is based on one
key observation: the timescale of the microscopic dynamics, which
for the rates (\ref{eq:hoprates}) is of order 1, is much faster than
that of the condensate motion. As shown below, the timescale of the
spilling process is of order $L$, validating this observation in the
thermodynamic limit. Due to this timescale separation, while the
condensate (i.e. the most occupied site), is static all other sites
reach a quasi-stationary distribution.

In both phases, by equating the LHS of \eqn (\ref{eq:mastereq}) to
zero the (quasi-)stationary distribution is found to be
\begin{equation}\label{eq:pofnsolution}
P_i(n) = P_i(0)\, z_i^n \prod_{k\leq n} \frac{1}{u(k)}, \qquad \text{with }
z_i \equiv J_i/\alphav{i}.
\end{equation}
Here, $z_i$ plays the role of a ``fugacity'' of site $i$. For rates
of the form (\ref{eq:hoprates}), the normalization of $P_i(n)$
yields
\begin{equation}\label{eq:p0}
P_i(0) = [{{}_2F_1(1,1;b+1;z_i)}]^{-1},
\end{equation}
where ${}_2F_1$ is a hypergeometric function (note that $P_i(0)$
depends on the exact form of $u(n)$ and not just on its large $n$
asymptotics). The occupation probability is asymptotically given by
$P_i(n) \sim n^{-b}z_i^{n}$.

We first examine the solution (\ref{eq:pofnsolution}) in the
subcritical and critical phases, and show that the model undergos a
condensation transition. Since the system is homogenous in these
phases, the subscript $i$ may be dropped from equations
(\ref{eq:betadef})--(\ref{eq:p0}). The fugacity can now be
determined in terms of the density by inverting the relation
\begin{equation}\label{eq:subcriticalrho}
\rho(z) = \sum_n nP(n) = \frac{{}_2F_1(2,2;b+2;z)}{(1+b)\,
{}_2F_1(1,1;b+1;z)}\,z,
\end{equation}
where the RHS is obtained by substituting \eqn
(\ref{eq:pofnsolution}) in the sum. Similarly, \eqn
(\ref{eq:betadef}) for $\alphavz$ reads in the homogeneous phases
$\alphavz = 1+ (\alpha - 1)[{}_2F_1(1,1;b+1;z)]^{-1}$.

The density (\ref{eq:subcriticalrho}) is an increasing function of
$z$ that attains its maximum at $z=1$, which is its radius of
convergence about the origin. A finite density at $z = 1$ indicates
a condensation phase transition, which is mathematically similar to
Bose-Einstein condensation \cite{evanszrpreview}. By substituting
$z=1$ in (\ref{eq:subcriticalrho}) it is seen that condensation
takes place when $b>2$, in which case the critical density is
$\rho_c = 1/(b-2)$, the same value as that of the usual ZRP. The
critical current is similarly found to be $J_c = \langle w
\rangle_{z\to 1}= 1 + (\alpha - 1)b/(b-1)$.
As long as $\rho < \rho_c$, the system remains in a homogeneous
subcritical phase. When $\rho$ is increased, the current $J$
increases until $\rho$ and $J$ reach their critical values and all
sites of the system are in a homogenous critical phase. When the
density is further increased, condensation sets in, breaking the
translational invariance of the system.

Let us now discuss the nonhomogeneous supercritical phase and the
mechanism of the condensate motion. We focus on the case of $\alpha
> 1$. In this case, the condensed phase is composed of a condensate,
which at any given time consists of two macroscopically occupied
consecutive sites (say 1 and 2), while the rest of the sites are
microscopically occupied. We show that $J_2 > 1$ and $J_i = 1$ for
$i \neq 2$. This results in an increase of the occupation of site 2
at the expense of site 1 over a macroscopic $O(L)$ time scale, while
the rest of the sites are in a quasi-stationary state. Therefore,
the condensate drifts with a velocity of order $L^{-1}$.

The analysis begins at site 1, whose occupation we assume is $n_1 =
O(L) \gg 1$, and thus it emits a mean current $J_2 =
\alphav{1}(1+\langle b/n_1 \rangle) \simeq \alphav{1}$. At the
moment, $\alphav{1}$ is unknown. It is determined self-consistently
at the end of the calculation. Since $P_L(0) \neq 0$, as is
established below, it is seen that $J_2 > 1$ (since $\alpha > 1$).
We now proceed to examine the second site. As long as site 1
accommodates the condensate it is never empty, i.e., $P_1(0) = 0$.
It follows from (\ref{eq:betadef}) that $\alphav{2}$ = 1. The
fugacity of the second site is then $z_2 \equiv J_2/\alphav{2}
\simeq J_2 > 1$, and therefore its occupation distribution
(\ref{eq:pofnsolution}) cannot be normalized. This means that as
long as site 1 is highly occupied, site 2 tends to accumulate
particles, implying that its occupation too becomes macroscopic (of
order $L$) for a long period of time
\cite{levine2005openzrp,Chertkov2010ZRPQueueing}. We call such a
site with fugacity $z_i > 1$ \emph{supercritical}.

The analysis now continues site by site in a similar fashion. For
each site $i$, $\alphav{i}$ is calculated using (\ref{eq:betadef})
from the known value of $P_{i-1}(0)$. The fugacity of site $i$
(\ref{eq:pofnsolution}) is then calculated from $\alphav{i}$ and the
incoming current into the site $J_i$. Once the fugacity is known,
$P_i(0)$ and $J_{i+1}$ are determined from (\ref{eq:p0}) and from
$J_{i+1} = \sum_n P_i(n) \alphav{i} u(n)$, and the process is
repeated in the next site. Performing this analysis reveals that
site 3 is critical (i.e., $z_3 = 1$) and sites $ i = 4,\ldots,L$ are
subcritical ($z_i < 1$), with $J_i = 1$ and $\alphav{i+1} =
1+(\alpha-1)/{{}_2F_1(1,1;b+1;1/\alphav{i})}$ for all $i\geq 3$.
This recursion relation defines a sequence $\alphav{i}$ which
converges (exponentially rapidly) to a unique fixed point
$w^*(\alpha,b)$ which is the solution of the equation
\begin{equation}\label{eq:betastar}
w^* = 1+\frac{\alpha-1}{{}_2F_1(1,1;b+1;1/w^*)},
\end{equation}
and thus satisfies $w^*(\alpha,b) > 1$ for all $\alpha > 1$. When
$L\gg1$, the periodic boundary conditions imply that $\alphav{1}
\simeq w^* > 1$, and thus \eqn (\ref{eq:pofnsolution}) confirms that
$P_L(0) > 0$. This closes the loop self-consistently and completes
the calculation of the quasi-stationary distribution for the
non-homogenous phase.

A natural order parameter for the condensation transition is the
bulk density of the ``background fluid'' $\rho_\text{BG}$, which can
be defined as the mean density of all but the two most occupied
sites (since the condensate is typically carried by two sites).
Below the transition, $\rho_\text{BG} = \rho$, which approaches
$\rho_c = 1/(b-2)$ as the transition is approached from below. Above
the transition, all sites outside of a finite boundary layer around
the condensate are subcritical with a mean occupation of
$\rho_\text{BG} \simeq {\rho(z=1/w^*)}<\rho(1) = \rho_c$ since the
function $\rho(z)$, \eqn (\ref{eq:subcriticalrho}), increases
monotonically with $z$. Therefore, the condensation transition is
found to be a discontinuous (first order) one. This is in contrast
to the usual ZRP with rates (\ref{eq:hoprates}) and $\alpha = 1$
where the transition is continuous. A similar discontinuity exists
in the current, which jumps from $J_c > 1$ just below $\rho_c$ to
$J=1$ just above it.

We now discuss the emergent dynamics of the condensate and identify
two distinct modes of motion: a regular slinky motion, and an
irregular motion through a barrier. The motion of the condensate
from one site to the next consists of two stages: a ``spilling''
stage during which it is supported on two sites, and a period before
this spilling is initiated, when the condensate is carried by a
single site. We first consider the spilling process. According to
the calculation above, the number of particles that accumulate in
the second condensate site per unit time is on average $J_2 - J_3 =
w^*-1$. As there are $N_\text{cond} \simeq (\rho-\rho_\text{BG})L$
particles in the condensate, the total spilling time
$T_\text{spill}$ scales, to leading order, linearly with the system
size: $T_\text{spill} = (\rho-\rho_\text{BG})L/(w^*-1)$. This
justifies the assumption of timescale separation which underlies the
existence of a quasi-stationary state. In the limit of $\alpha \to
1$, the spilling time diverges.

Once a spilling is complete, there is a moment that the condensate
is located solely on a single site. We now relabel this site as site
1. At this moment, the occupation of the following site is $n_2
\approx \rho_\text{BG} = O(1)$. The rate at which particles leave
the second site is, at this stage, approximately $J_3 \approx
1+b/\rho_\text{BG}$, which should be compared with the rate of
incoming particles, $J_2 \simeq w^*$. According to \eqn
(\ref{eq:betastar}) and the definition of $\rho_\text{BG}$, the two
rates are equal when $\alpha = \alpha^*$ which is the solution of
the equation $\alpha^* =
1+b\,{}_2F_1\bigl(1,1;b+1;1/w^*(b,\alpha^*)\bigr)/\rho_\text{BG}(b,\alpha^*)$.
The mode of condensate motion now depends on whether $\alpha$ is
larger or smaller than $\alpha^*$. (i) When $\alpha > \alpha^*$, the
initial current into site 2 is larger than the mean current out of
this site, and a spilling of the condensate is initiated
immediately. In this case, the condensate drifts continuously in a
slinky motion as in \fig \ref{fig:CondDrift}b. (ii) On the other
hand, $J_2<J_3$ when $1 < \alpha < \alpha^*$, and thus particles do
not immediately accumulate on site 2. The incoming current into the
site surpasses the outgoing current and spilling sets in only after
fluctuations bring the occupation of the second site to a value
$n^*(\alpha,b)$ which is defined by $w^* = 1+b/n^*$. The ensuing
motion of the condensate is more irregular, with a stable condensate
which occasionally spills to the next site as in \fig
\ref{fig:CondDrift}c. Note that the mean time $T_\text{barrier}$ it
takes before a fluctuation brings $n_2$ over the barrier $n^*$ does
not scale with the system size. Thus, in a large enough system the
condensate regularly drifts and is typically supported by two
neighboring sites for any value of $\alpha > 1$.

\begin{figure}
        \includegraphics[width=0.35\textwidth]{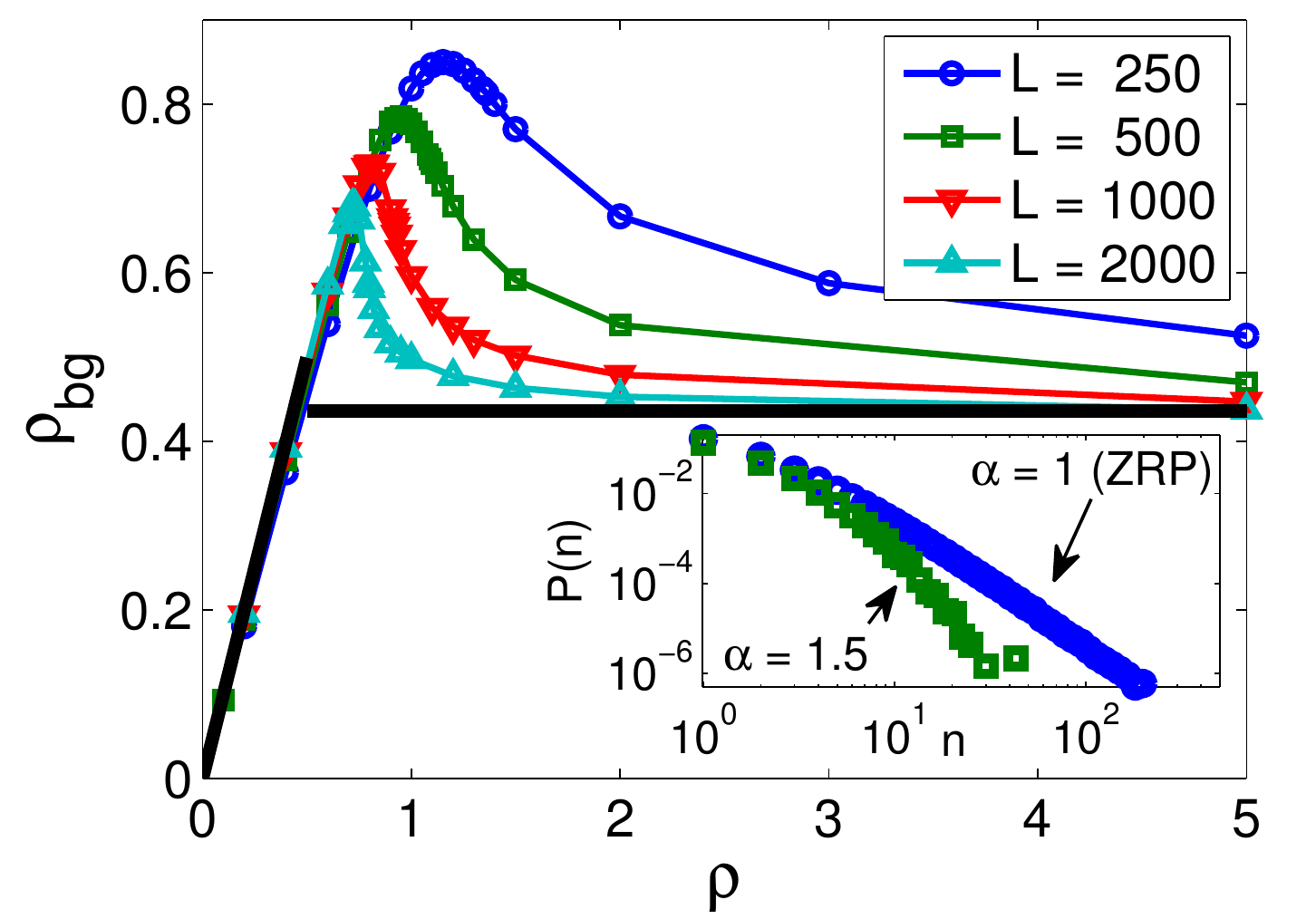}
        \caption{A first order phase transition is seen in the background
        density $\rho_\text{BG}$ as a function of the density $\rho$.
        Numerical results for several system sizes are plotted, along with
        an extrapolation to $L = \infty$ (thick black line).
        Here $\alpha = 1.5$ and $b = 3$. The inset
        shows that the occupation of a site far away from the condensate
        has a subcritical distribution (i.e.,
        with an exponential tail) when $\alpha > 1$.
        This differs from the usual
        ZRP where the background fluid is known to be critical.
        Results are for $L=1000$, $\rho = 10$, and $b=3$.
        \label{fig:fig2}}
\end{figure}

The numerical simulations support the qualitative picture which
emerges from the MF analysis presented above, even if the two differ
quantitatively. The existence of two modes of motion, slinky-like
and irregular, and the crossover between them as $\alpha$ is
increased conform with numerical findings (\fig
\ref{fig:CondDrift}). In particular, the spilling mechanism between
the two condensate sites in which the accumulation of particles is
linear in time is verified (\fig \ref{fig:CondDrift}b). Furthermore,
the first order nature of the transition, as manifested by the
$\rho_\text{bg}(\rho)$ curve, and the subcritical nature of the
background fluid are presented in \fig \ref{fig:fig2}.

The mechanism for the condensate drift found in this model can be
summarized as follows: the spontaneous breaking of translation
invariance by the formation of the condensate may induce an
accumulation of particles in a nearby site. This accumulation
results in a continually drifting condensate, since whenever a
condensate is established on a new site, another one begins to form
further ahead. This mechanism holds in a much more general setting,
including when other forms $w(n)$, partially asymmetric hopping and
higher dimensional lattices are considered, and more widely in other
non-factorized mass-transport models \cite{FutureLongPaper}. Note
that the drift discussed here, in which the two most occupied sites
are typically nearest neighbors, cannot occur in models with a
factorized steady state, the latter being symmetric under site
permutations. In this respect, our mechanism differs from that
studied recently in \cite{BartekEvans2012Explosive}, where unbounded
hopping rates generate a drift (with infinite velocity) in a model
whose steady state factorizes.

An important point to note is that in general, the new condensate
site does not have to be a neighbor of the old one. For instance, in
our model (\ref{eq:modeldscrp})--(\ref{eq:hoprates}) with $\alpha <
1$, a similar analysis shows that the condensate skips every other
site, as observed in \fig \ref{fig:CondDrift}d
\cite{FutureLongPaper}. In this case, the supercritical site is site
3, rather than 2 (when the condensate is located on site 1). In
principle, it may happen that there is more than one supercritical
site, possibly leading to more complicated condensate drifts. It may
also happen that no other site is supercritical, in which case the
condensate would not drift. A precise and general classification of
the conditions under which a condensate drift occurs remains an
interesting open problem. However, in many specific models, a study
of condensation and the condensate motion can be carried out
following the mean-field procedure outlined in this Letter. For
instance, a recently proposed accelerated exclusion process (AEP)
\cite{dong2012AEP} can be analyzed in a similar fashion, yielding
the phase diagram of the model and revealing that the AEP condensate
drifts in the steady-state \cite{FutureAEP}.

\begin{acknowledgments}
We thank Ariel Amir, Amir Bar, Or Cohen and Tridib Sadhu for useful
discussions. The support of the Israel Science Foundation (ISF) is
gratefully acknowledged.
\end{acknowledgments}


\bibliographystyle{h-physrev}
\bibliography{cond_drift_arxiv}

\end{document}